\pdfoutput=1

\documentclass{hotnets17}

\usepackage{times}  
\usepackage[hidelinks]{hyperref}

\hypersetup{pdfstartview=FitH,pdfpagelayout=SinglePage}

\usepackage{mathptmx} 
\usepackage{amsmath}
\usepackage{amssymb}
\usepackage{amsfonts}
\usepackage{bbding}

\usepackage[font=bf]{caption}
\usepackage{quoting}
\usepackage{csquotes}
\SetBlockEnvironment{quoting}

\usepackage{subfig}

\usepackage{multirow}
\usepackage[linesnumbered,noend,boxed]{algorithm2e}
\newcommand{\subparagraph}{} 
\usepackage[usenames,dvipsnames]{color}

\usepackage[compact]{titlesec}

\usepackage{amsmath}

\usepackage{times}

\usepackage{xspace}
\usepackage{epsfig} 
\usepackage{amsmath}

\usepackage{booktabs}
\usepackage{multirow}
\usepackage{graphicx}
\usepackage[table,xcdraw]{xcolor}

\usepackage{listings}
\usepackage{fancyvrb}
\usepackage{mathrsfs}
\usepackage[bordercolor=white,backgroundcolor=gray!30,linecolor=black,colorinlistoftodos]{todonotes}
\VerbatimFootnotes

\usepackage[12h=false]{scrtime}

\usepackage{tablefootnote}

\usepackage[normalem]{ulem}

\usepackage{titlesec}
\usepackage{etoolbox}
\usepackage{url}

\makeatletter
\patchcmd{\ttlh@hang}{\parindent\z@}{\parindent\z@\leavevmode}{}{}
\patchcmd{\ttlh@hang}{\noindent}{}{}{}
\makeatother

\usepackage[protrusion=true,expansion=true]{microtype}
\newtheorem{theo}{Insight}

\newcounter{insightlabel}
\newcounter{insightnmbr}
\renewcommand{\theinsightlabel}{{\theinsightnmbr}}

\newcounter{implabel}
\newcounter{impnmbr}
\renewcommand{\theimplabel}{{\theimpnmbr}}

\newcommand{\tightcaption}[1]{\vspace{-0.12cm}\caption{{\em #1}}\vspace{-0.13cm}}

\newcommand{\eg}{{\it e.g.,}\xspace}
\newcommand{\ie}{{\it i.e.,}\xspace}

\newcommand{\comment}[1]{}
\newcounter{note}[section]

\usepackage{pifont}

\newcommand{\mypara}[1]{\smallskip\noindent{\bf {#1}:}~}

\newcommand{\sid}[1]{{\footnotesize\color{orange}{}}} 
 
\newcommand{\xil}[1]{{\footnotesize\color{magenta}{}}}
\newcommand{\hm}[1]{{\footnotesize\color{Maroon}{}}}
\newcommand{\ai}[1]{{\footnotesize\color{Purple}{}}}

\newcommand{\cameraremove}[1]{{\color{black}{}}\xspace}

\newcommand{\vyasremove}[1]{{\color{red}{}}\xspace}

\newcommand{\old}[1]{{\color{black}{}}}

\newcounter{packednmbr}

\newenvironment{packedenumerate}{\begin{list}{\thepackednmbr.}{\usecounter{packednmbr}\setlength{\itemsep}{0.5pt}\addtolength{\labelwidth}{-4pt}\setlength{\leftmargin}{\labelwidth}\setlength{\listparindent}{\parindent}\setlength{\parsep}{1pt}\setlength{\topsep}{0pt}}}{\end{list}}

\newenvironment{packeditemize}{\begin{list}{$\bullet$}{\setlength{\itemsep}{0.5pt}\addtolength{\labelwidth}{-4pt}\setlength{\leftmargin}{\labelwidth}\setlength{\listparindent}{\parindent}\setlength{\parsep}{1pt}\setlength{\topsep}{0pt}}}{\end{list}}

\begin{document}



\title{Scaling Video Analytics Systems \\ to Large Camera Deployments}
\author{
	\Large Samvit Jain\\
	\large University of California, Berkeley\\
	\normalsize samvit@eecs.berkeley.edu
	\and
	\Large Ganesh Ananthanarayanan\\
	\large Microsoft Research\\
	\normalsize ga@microsoft.com
	\and
	\Large Junchen Jiang\\
	\large University of Chicago\\
	\normalsize junchenj@uchicago.edu
	\and \\[-6pt]
	\Large Yuanchao Shu\\
	\large Microsoft Research\\
	\normalsize yuanchao.shu@microsoft.com
	\and \\[-6pt]
	\Large Joseph Gonzalez\\
	\large University of California, Berkeley\\
	\normalsize jegonzal@cs.berkeley.edu
}

\maketitle

\begin{abstract}

Driven by advances in computer vision and the falling costs of camera hardware, organizations are deploying video cameras {\em en masse} for the spatial monitoring of their physical premises.
Scaling video analytics to massive camera deployments, however, presents a new and mounting challenge, as compute cost grows proportionally to the number of camera feeds.
This paper is driven by a simple question: can we scale video analytics in such a way that {\em cost grows sublinearly}, or even remains constant, as we deploy more cameras, while {\em inference accuracy remains stable}, or even improves.
We believe the answer is yes. Our key observation is that video feeds from wide-area camera deployments demonstrate significant content correlations (e.g. to other geographically proximate feeds), both in space and over time.
These \textit{spatio-temporal correlations} can be harnessed to dramatically reduce the size of the  inference search space, decreasing both workload and false positive rates in multi-camera video analytics.
By discussing use-cases and technical challenges, we propose a roadmap for scaling video analytics to large camera networks, and outline a plan for its realization.

\end{abstract}


\section{Introduction}

Driven by plummeting camera prices and the recent successes of computer vision-based video inference, organizations are deploying cameras at scale for applications ranging from surveillance and flow control to retail planning and sports broadcasting \cite{ristani2018, optasia, prw2017}.
Processing video feeds from large deployments, however, requires a proportional investment in either compute hardware (i.e. expensive GPUs) or cloud resources (i.e. GPU machine time), costs from which easily exceed that of the camera hardware itself \cite{wyzecam, TeslaP100, EC2Pricing}.
A key reason for these large resource requirements is the fact that, today, video streams are analyzed {\em in isolation}. 
As a result, the compute required to process the video grows linearly with the number of cameras. 
We believe there is an opportunity to both stem this trend of linearly increasing costs, \textit{and} improve accuracy, by viewing the cameras {\em collectively}.

\begin{figure}[t!]
    \centering
    \includegraphics[width=0.5\textwidth]{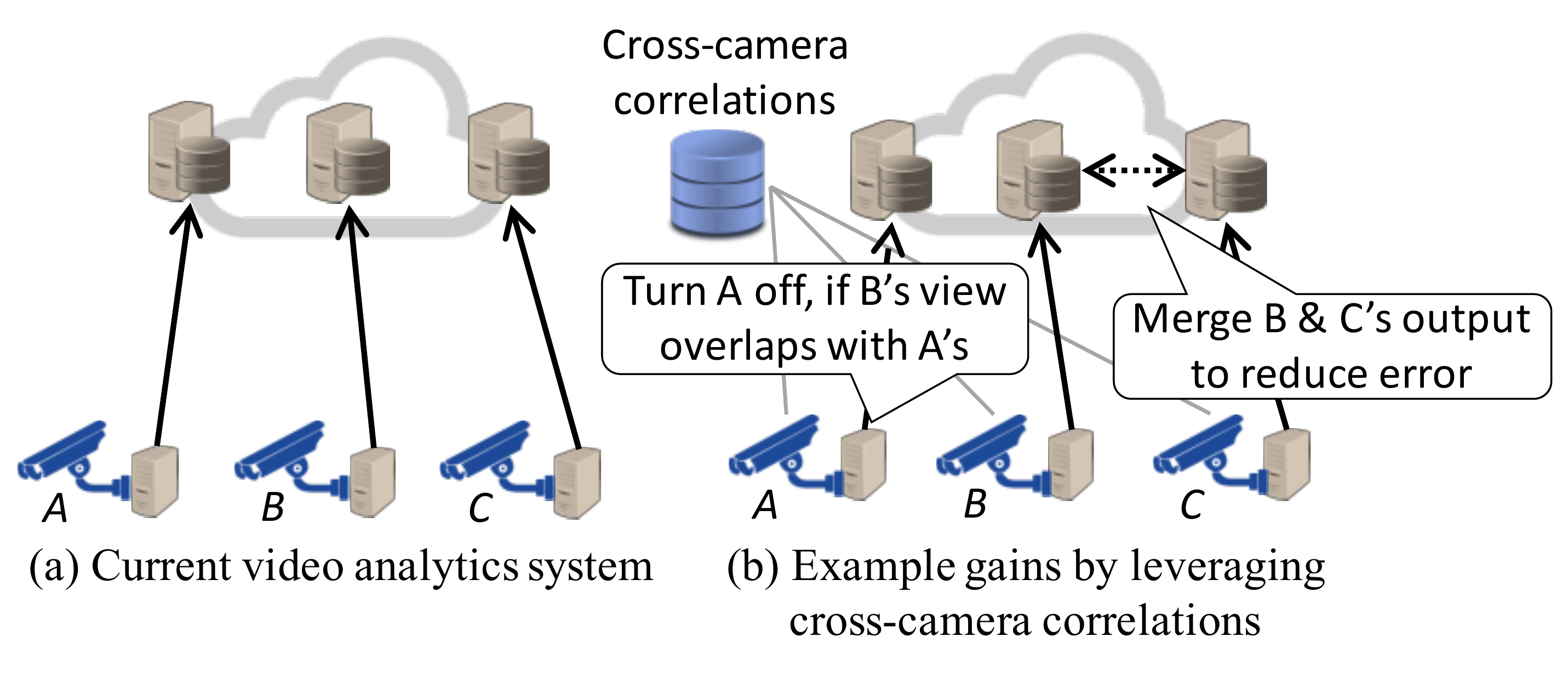}
    \vspace{-6mm}
    \tightcaption{Contrasting (a) traditional per-camera video analytics with (b) the proposed approach that leverages 
    cross-camera correlations.}
	\vspace{+2mm}
    \label{fig:architecture}
\end{figure}

This position paper is based on a simple observation---cameras deployed over any geographic area, whether a large campus, an outdoor park, or a subway station network, demonstrate significant content correlations---both spatial and temporal.
For example, nearby cameras may perceive the same objects, though from different angles or at different points in time.
We argue that these cross-camera correlations can be harnessed, so as to use {\em substantially fewer resources} and/or achieve {\em higher inference accuracy} than a system that runs complex inference on all video feeds independently.
For example, if a query person is identified in one camera feed, we can then exclude the possibility of the individual appearing in a distant camera within a short time period.
This eliminates extraneous processing and reduces the rate of false positive detections (Figure~\ref{fig:architecture}(a)).
Similarly, one can improve accuracy by combining the inference results of multiple cameras that monitor the same objects from different angles (Figure~\ref{fig:architecture}(b)).
Our initial evaluation on a real-world dataset with eight cameras shows that using cameras collectively can yield resource savings of {\em at least 74\%}, while also improving inference accuracy.
More such opportunities are outlined in \S\ref{sec:benefits}.

Given the recent increase in interest in systems infrastructure for video analytics \cite{videostorm, mcdnn, chameleon}, we believe the important next step for the community is designing a software stack for \textit{collective} camera analytics.
Video processing systems today generally analyze video streams independently even while useful cross-camera correlations exist \cite{videostorm, noscope, chameleon}.
On the computer vision side, recent work has tackled specific multi-camera applications (\eg tracking~\cite{ristani2018, prw2017, song2010tracking}), but has generally neglected the growing cost of inference itself.

We argue that the key to scaling video analytics to large camera deployments lies in fully leveraging these latent cross-camera correlations. 
We identify several architectural aspects that are critical to improving resource efficiency and accuracy but are missing in current video analytics systems.
First, we illustrate the need for a new system module that learns and maintains up-to-date {\em spatio-temporal correlations} across cameras.
Second, we discuss online pipeline reconfiguration and composition, where video pipelines incorporate information from other correlated cameras (\eg to eliminate redundant inference, or ensemble predictions) to save on cost or improve accuracy.
Finally, to recover missed detections arising from our proposed optimizations, we note the need to process small segments of historical video at faster-than-real-time rates, alongside analytics on live video.

Our goal is not to provide a specific system implementation, but to motivate the design of an accurate, cost-efficient {\em multi-camera} video analytics system.
In the remainder of the paper, we will discuss the trends driving the proliferation of large-scale analytics (\S\ref{sec:trends}), enumerate the advantages of using cross-camera correlations (\S\ref{sec:benefits}), and outline potential approaches to building such a system (\S\ref{sec:architecture}).
We hope to inspire the practical realization of these ideas in the near future.

\section{Camera trends \& applications}
\label{sec:trends}

This section sets the context for using many cameras collaboratively by discussing (1) trends in camera deployments, and (2) the recent increase in interest in cross-camera applications. 

\mypara{Dramatic rise in smart camera installations}
Organizations are deploying cameras {\em en masse} to cover physical areas. Enterprises are fitting cameras in office hallways, store aisles, and building entry/exit points; government agencies are deploying cameras outdoors for surveillance and planning. Two factors are contributing to this trend:

\begin{packedenumerate}

\item {\em Falling camera costs} enable more enterprises and business owners to install cameras, and at higher density. 
For instance, today, one can install an HDTV-quality camera with on-board SD card storage for \$20~\cite{wyzecam}, whereas three years ago the industry's first HDTV camera cost \$1,500~\cite{infowatch16}.

\item {\em Falling camera costs} allow more enterprises and business owners to install cameras, and at higher density. 
For instance, today, one can install an HDTV-quality camera with on-board SD card storage for \$20 ~\cite{wyzecam}, where as three years ago the industry's first HDTV camera cost \$1,500~\cite{infowatch12}.
Driven by the sharp drop in camera costs, camera installations have grown exponentially, with 566~PT of data generated by {\em new} video surveillance cameras worldwide {\em every day} in 2015, compared to 413~PT generated by newly installed cameras in 2013~\cite{infowatch16}.

There has been a recent wave of interest in ``AI cameras'' -- cameras with compute and storage on-board -- that are designed for processing and storing the videos~\cite{awsdeeplens, google-clip, qualcommvip}. These cameras are programmable and allow for running arbitrary deep learning models as well as classic computer vision algorithms. AI cameras are slated to be deployed at mass scales by enterprises.

\item {\em Advances in computer vision}, specifically in object detection and re-identification techniques~\cite{prw2017,Yolo2016}, have sparked renewed interest among organizations in camera-based data analytics. For example, transportation departments in the US are moving to use video analytics for traffic efficiency and planning~\cite{VisionZero2017}. A key advantage of using cameras is that they are relatively easy to deploy and can be purposed for multiple objectives.

\end{packedenumerate}

\mypara{Increased interest in cross-camera applications}
We focus on applications that involve video analytics {\em across} cameras.
While many cross-camera video applications were envisaged in prior research, the lack of one or both of the above trends made them either prohibitively expensive or insufficiently accurate for real-world use-cases.
 
We focus on a category of applications we refer to as {\em spotlight search}.
Spotlight search refers to detecting a specific type of activity and object (\eg shoplifting, a person), and then tracking the entity as it moves through the camera network.
Both detecting activities/objects and tracking require compute-intensive techniques, \eg face recognition and person re-identification~\cite{prw2017}.
Note that objects can be tracked both in the forward direction (``real-time tracking''), and in the backward direction (``investigative search'') on recorded video.
Spotlight search represents a broad template, or a core building block, for many cross-camera applications. Cameras in a retail store use spotlight search to monitor customers flagged for suspicious activity. Likewise, traffic cameras use spotlight search to track vehicles exhibiting erratic driving patterns.
In this paper, we focus on spotlight search on live camera feeds as the canonical cross-camera application.

\mypara{Metrics of interest}
The two metrics of interest in video analytics applications are inference {\em accuracy} and {\em cost} of processing. Inference accuracy is a function of the model used for the analytics, the labeled data used for training, and video characteristics such as frame resolution and frame rate \cite{videostorm, chameleon, noscope}. All of the above metrics also influence the {\em cost} of processing -- larger models and higher quality videos enable higher accuracy, at the price of increased resource consumption or higher processing latency.
When the video feeds are analyzed at an edge or cloud cluster, cost also includes the bandwidth cost of sending the videos over a wireless network, which increases with the number of video feeds.

\section{New opportunities in camera deployments}
\label{sec:benefits}

Next, we explain the key benefits -- in efficiency and accuracy -- of cross-camera video analytics. 
The key insight is that scaling video analytics to many cameras does not necessarily stipulate a linear increase in cost; 
instead, one can significantly improve cost-efficiency as well as accuracy by leveraging the spatio-temporal 
correlations across cameras.

\subsection{Key enabler: Cross-camera correlations}
\label{subsec:correlations}
A fundamental building block in enabling cross-camera collaboration are the profiles of {\em spatio-temporal correlations} across cameras.
At a high level, these spatio-temporal correlations capture the relationship between the content of camera $A$ and the content of camera $B$ over a time delta $\Delta t$.\footnote{The correlation reduces to ``spatial-only'', when $\Delta t \rightarrow 0$.}
This correlation manifests itself in at least three different forms.
First, the same object can appear in multiple cameras, \ie~{\em content} correlation, at the same time (\eg cameras in the same room) or at different points in time (\eg cameras placed at two ends of a hallway).
Second, multiple cameras may share similar characteristics, \ie~{\em property} correlation, \eg the types, velocities, and sizes of contained objects.
Third, one camera may have a different viewpoint on objects than another, resulting in a {\em position} correlation, \eg some cameras see larger/clearer faces since they are deployed closer to eye level.

As we will show next, the prevalence of these cross-camera correlations in dense camera networks enables key opportunities to use the compute (CPU, GPU) and storage (RAM, SSD) resources on these cameras {\em collaboratively}, by leveraging their network connectivity.

\vspace{0.1cm}
\subsection{Better cost efficiency}
\label{subsec:cost}

\begin{figure}[t!]
	\centering
	\includegraphics[width=0.48\textwidth]{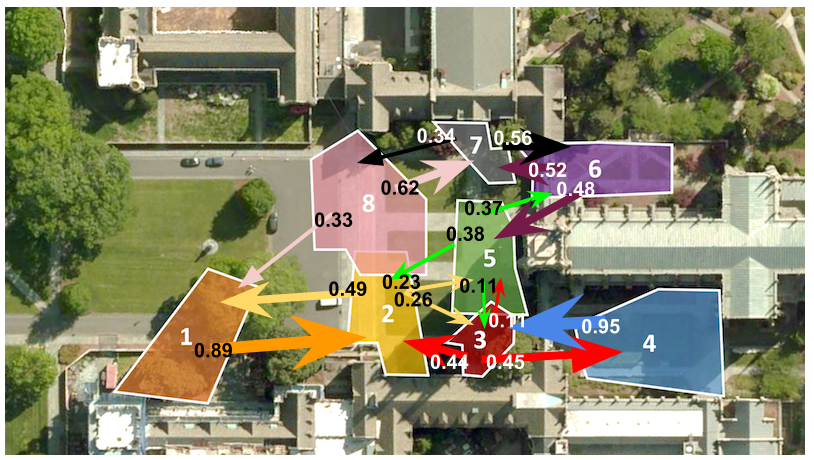}
	\vspace{-5mm}
	\tightcaption{Camera topology and traffic flow (in \% of outgoing traffic) in the DukeMTMC dataset.}
	\label{fig:dukemtmc_map}
	\vspace{+3mm}
\end{figure}

Leveraging cross-camera correlations improves the {\em cost efficiency} of multi-camera video analytics, without adversely impacting accuracy. Here are two examples.

\vspace{0.1cm}
\noindent{\bf C1: Eliminating redundant inference}

In cross-camera applications like spotlight search, there are often far fewer objects of interest than cameras.
Hence, ideally, query resource consumption over multiple cameras should not grow proportionally to the number of cameras.
We envision two potential ways of doing this by leveraging content-level correlations across cameras (\S\ref{subsec:correlations}).

\begin{packeditemize}
	
\item When two spatially correlated cameras have overlapping views (\eg cameras covering the same room or hallway), the overlapped region need only be analyzed once.

\item When an object leaves a camera, only a small set of relevant cameras (\eg cameras likely to see the object in the next few seconds), identified via their spatio-temporal correlation profiles, need search for the object.

\end{packeditemize}

In spotlight search, for example, once a suspicious activity or individual is detected, we can selectively trigger multi-class object detection or person re-identification models only on the cameras that the individual is likely to traverse.
In other words, we can use spatio-temporal correlations to narrow the search space by \textit{forecasting} the trajectory of objects.

We analyze the popular ``DukeMTMC'' video dataset~\cite{ristani2016MTMC}, which contains footage from eight cameras on the Duke University campus.
Figure~\ref{fig:dukemtmc_map} shows a map of the different cameras, along with the percentage of traffic leaving a particular camera $i$ that next appears in another camera $j$.
Figures are calculated based on manually annotated human identity labels.
As an example observation, within a time window of $90$ minutes, $89\%$ of all traffic leaving Camera 1 first appears at Camera 2.
At Camera 3, an equal percentage of traffic, about $45\%$, leaves for Cameras 2 and 4.
Gains achieved by leveraging these spatial traffic patterns are discussed in \S\ref{subsec:results}.

\vspace{0.1cm}
\noindent{\bf C2: Resource pooling across cameras}

Since objects/activities of interest are usually sparse, most cameras do not need to run analytics models all the time. This creates a substantial heterogeneity in workloads across different cameras.
For instance, one camera may monitor a central hallway and detect many candidate persons, while another camera detects no people in the same time window.

\begin{figure}[t!]
    \centering
    \includegraphics[width=0.45\textwidth]{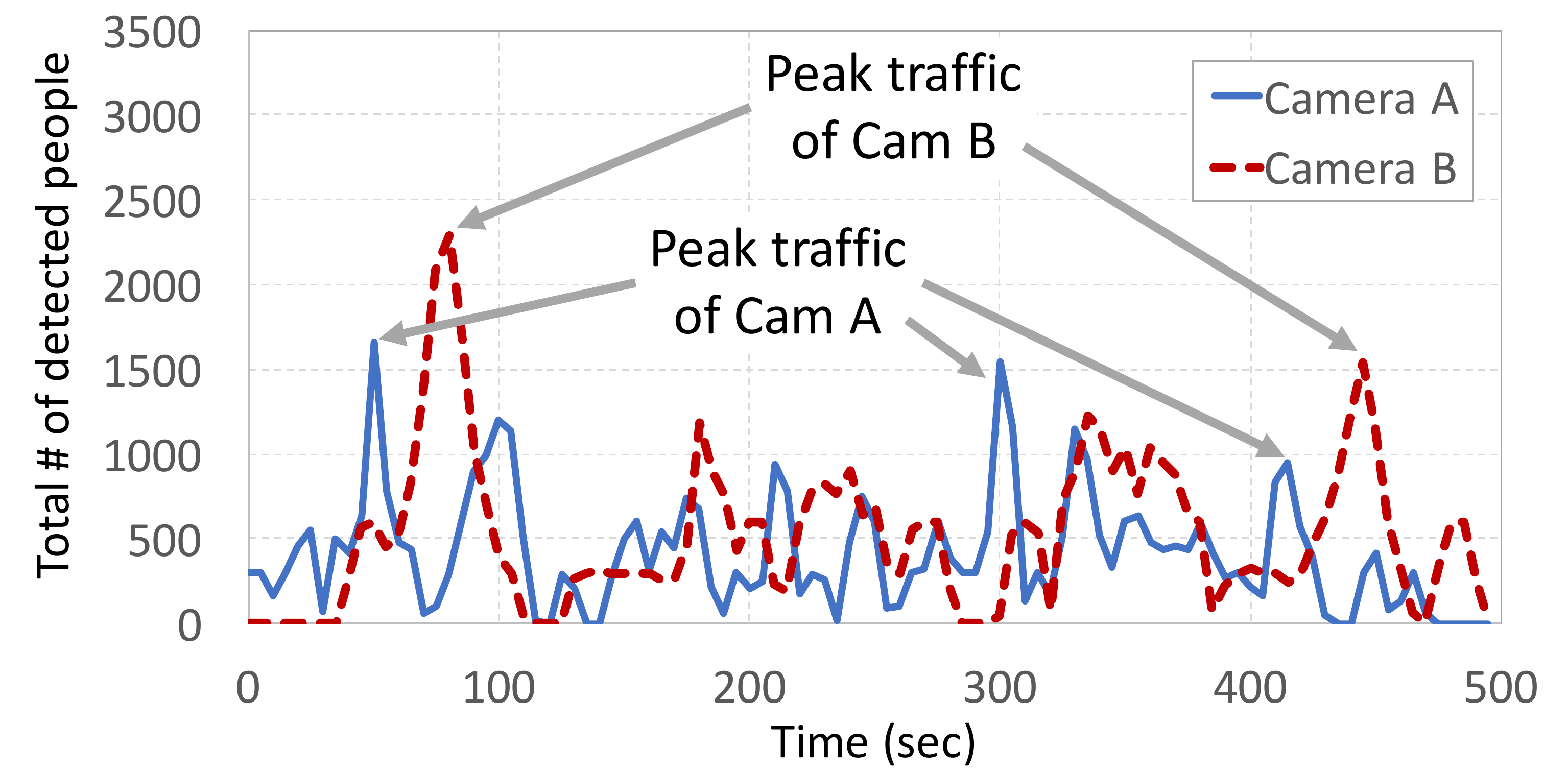}
    \vspace{-1mm}
    \tightcaption{Number of people detections on different cameras in the DukeMTMC dataset.}
    \label{fig:dukemtmc_imbalance}
    \vspace{+2mm}
\end{figure}

Such workload heterogeneity provides an opportunity for dynamic offloading, in which more heavily utilized cameras transfer part of their analytics load to less-utilized cameras.
For instance, a camera that runs complex per-frame inference can offload queries on some frames to other ``idle'' cameras whose video feeds are static.
Figure~\ref{fig:dukemtmc_imbalance} shows the evident imbalance in the number of people detected on two cameras across a 500 second interval. 
By {\em pooling} available resources and balancing the workload across multiple cameras, one can greatly reduce resource provisioning on each camera (\eg deploy smaller, cheaper GPUs), from an allocation that would support peak workloads.
Such a scheme could also reduce the need to stream compute tasks to the cloud, a capability constrained by available bandwidth and privacy concerns.
{\em Content} correlations, \S\ref{subsec:correlations} directly facilitate this offloading as they foretell query trajectories, and by extension, workload.

\subsection{Higher inference accuracy}
\label{subsec:accuracy}
We also observe opportunities to improve inference accuracy, without increasing resource usage.

\vspace{0.1cm}
\noindent{\bf A1: Collaborative inference}

Using an ensemble of identical models to render a prediction is an established method for boosting inference accuracy~\cite{Hinton2015}. The technique also applies to model ensembles consisting of multiple, correlated video pipelines (\eg with different perspectives on an object). 
Inference can also benefit from hosting dissimilar models on different cameras. For instance, camera A with limited resources uses a specialized, low cost model for flagging cars, while camera B uses a general model for detection. Then camera A can offload its video to camera B to cross-validate its results when B is idle.

Cameras can also be correlated in a {\em mutually exclusive} manner.
In spotlight search, for example, if a person $p$ is identified in camera $A$, we can preclude a detection of the same $p$ in another camera whose view does not overlap with $A$.
Knowing where an object is likely {\em not} to show up can significantly improve tracking precision over a na\"ive baseline that searches all of the cameras.
In particular, removing unlikely candidates from the space of potential matches reduces false positive matches, which tend to dislodge subsequent tracking and bring down precision (see \S\ref{subsec:results}).

\vspace{0.1cm}
\noindent{\bf A2: Cross-camera model refinement}

One source of video analytics error stems from the fact that objects look differently in real-world settings than in training data.
For example, some surveillance cameras are installed on ceilings, which reduces facial recognition accuracy, due to the oblique viewing angle~\cite{npr-face}.
These errors can be alleviated by retraining the analytics model, using the output of another camera that has an eye-level view as the ``ground truth''.
As another example, a traffic camera under direct sunlight or strong shadows will tend to render poorly exposed images, resulting in lower detection and classification accuracy than images from a camera without lighting interference~\cite{fu17}. Since lighting conditions change over time, two such cameras can complement each other, via collaborative model training.
Opportunities for such cross-camera model refinement are a direct implication of position correlations (\S\ref{subsec:correlations}) across cameras.

\subsection{Preliminary results}
\label{subsec:results}

Table \ref{tbl:spotlight-results} contains a preliminary evaluation of our spotlight search scheme on the Duke dataset~\cite{ristani2016MTMC}, which consists of 8 cameras.
We quantify resource savings by computing the ratio of (a) the number of person detections processed by the baseline (i.e. 76,500) to (b) the number of person detections processed by a particular filtering scheme (e.g. 22,500).
Observe that applying spatio-temporal filtering results in significant resource savings and much higher precision, compared to the baseline, at the price of slightly lower recall.

\begin{table}[t]
	\tightcaption{Spotlight search results for various levels of spatio-temporal correlation filtering. A filtering level of $k\%$ signifies that a camera must receive $k\%$ of the traffic from a particular source camera to be searched. Larger $k$ (e.g. $k=10$) corresponds to more aggressive filtering, while $k=0$ corresponds to the baseline, which searches all of the cameras. All results are reported as aggregate figures over 100 tracking queries on the 8 camera DukeMTMC dataset \protect\cite{ristani2016MTMC}.}
	\centering
	\label{tbl:spotlight-results}
	\begin{tabular}{@{\extracolsep{4pt}}lcccc}
		\toprule
		Filtering & Detections & Savings & Recall & Precis.\\
		level (\%) & processed & (vs. baseline) & (\%) & (\%) \\
		\midrule
		0\% & 76,510 & 0.0 & \textbf{57.4} & 60.6 \\
		1\% & 29,940 & 60.9 & 55.0 & 81.4 \\
		3\% & 22,490 & 70.6 & 55.1 & 81.9 \\
		10\% & 19,639 & \textbf{74.3} & 55.1 & \textbf{81.9} \\
		\bottomrule
	\end{tabular}
	\vspace{+2mm}
\end{table}


\section{Architecting for cross-camera analytics}
\label{sec:architecture}

We have seen that exploiting spatio-temporal correlations across cameras can improve cost efficiency and inference accuracy in multi-camera settings. 
Realizing these benefits in practice, however, requires re-architecting the underlying video analytics stack. 
This section articulates the key missing pieces in current video analytics systems, and outlines the core technical challenges that must be addressed in order to realize the benefits of collaborative analytics.

\subsection{What's missing in today's video analytics?}
\label{subsec:missing}

The proposals in \S\ref{subsec:cost} and \S\ref{subsec:accuracy} require four basic capabilities.

\mypara{\#1: Cross-camera correlation database}
First, a new system module must be introduced to learn and maintain an up-to-date view of the spatio-temporal correlations between any pair of cameras (\S\ref{subsec:correlations}). 
Physically, this module can be a centralized service, or a decentralized system, with each camera maintaining a local copy of the correlations. 
Different correlations can be represented in various ways.
For example, content correlations can be modeled as the \textit{conditional probability} of detecting a specific object in camera B at time $t$, given its appearance at time $t - \Delta t$ in camera A, and stored as a discrete, 3-D matrix in a database.
This database of cross-camera correlations must be dynamically updated, because the correlations between cameras can vary over time: video patterns can evolve, cameras can enter or leave the system, and camera positions and viewpoints can change.
We discuss the intricacy of discovering these correlations, and the implementation of this new module, in \S\ref{subsec:challenges}.

\begin{figure}[t!]
    \centering
    \includegraphics[width=0.41\textwidth]{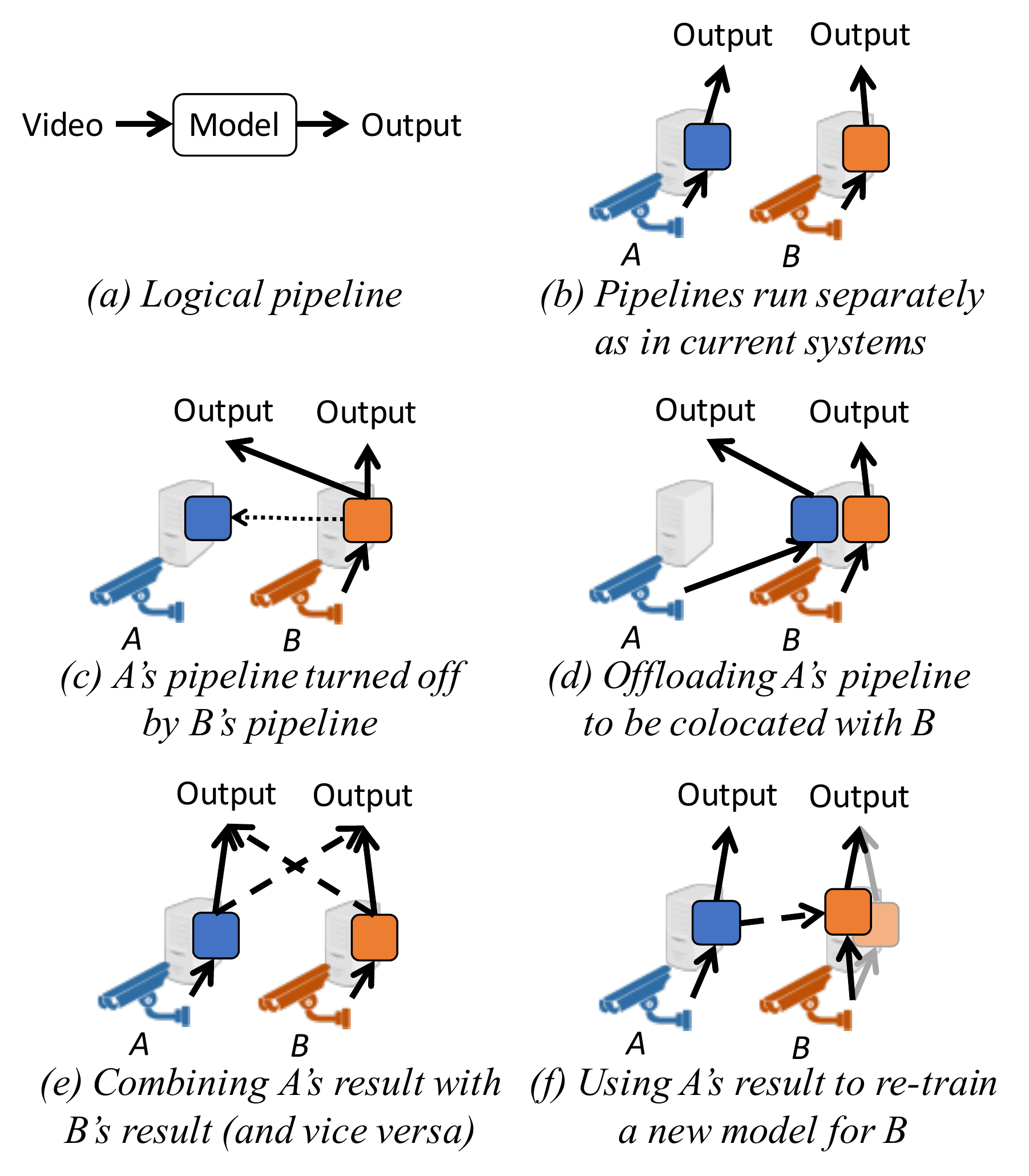}
    \vspace{-0.1cm}
    \tightcaption{Illustrative examples of peer-triggered reconfiguration (c, d) and pipeline composition (e, f) using an example logical pipeline (a) running on two cameras (b).}
    \vspace{+0.2cm}
    \label{fig:missing-pieces}
\end{figure}

\mypara{\#2: Peer-triggered inference}
Today, the execution of a video analytics pipeline (what resources to use and which video to analyze) is largely pre-configured.
To take advantage of cross-camera correlations, an analytics pipeline must be aware of the inference results of other relevant video streams, and support {\em peer-triggered inference} at runtime.
Depending on the content of other related video streams, an analytics task can be assigned to the compute resources of {\em any} relevant camera to process {\em any} video stream at {\em any} time. 
This effectively {\em separates} the logical analytics pipeline from its execution.
To eliminate redundant inference (C1 of \S\ref{subsec:cost}), for instance, one video stream pipeline may need to dynamically trigger (or switch off) another video pipeline (Figure~\ref{fig:missing-pieces}.c).
Similarly, to pool resources across cameras (C2 of \S\ref{subsec:cost}), a video stream may need to dynamically offload computation to another camera, depending on correlation-based workload projections (Figure~\ref{fig:missing-pieces}.d).
To trigger such inference, the current inference results need to be shared in real-time \textit{between} pipelines.
While prior work explores task offloading across cameras and between the edge and the cloud~\cite{Cuervo2010,LiKamWa2013}, the trigger is usually workload changes on a single camera.
We argue that such dynamic triggering must also consider events on the video streams of other, related cameras.

\mypara{\#3: Video pipeline composition}
Analyzing each video stream in isolation also precludes learning from the content of other camera feeds.
As we noted in \S\ref{subsec:accuracy}, by combining the inference results of multiple correlated cameras, \ie composing multiple video pipelines, one can significantly improve inference accuracy.
Figure~\ref{fig:missing-pieces} shows two examples.
Firstly, by sharing inference results across pipelines in real-time (Figure~\ref{fig:missing-pieces}.e), one can correct the inference error of another less well-positioned camera (A1 in \S\ref{subsec:accuracy}). 
Secondly, the inference model for one pipeline can be refined/retrained (Figure~\ref{fig:missing-pieces}.f) based on the inference results of another better positioned camera (A2 in \S\ref{subsec:accuracy}).
Unlike the aforementioned reconfiguration of video pipelines, \textit{merging} pipelines in this way actually impacts inference output.

\begin{figure}[t!]
    \centering
    \includegraphics[width=0.5\textwidth]{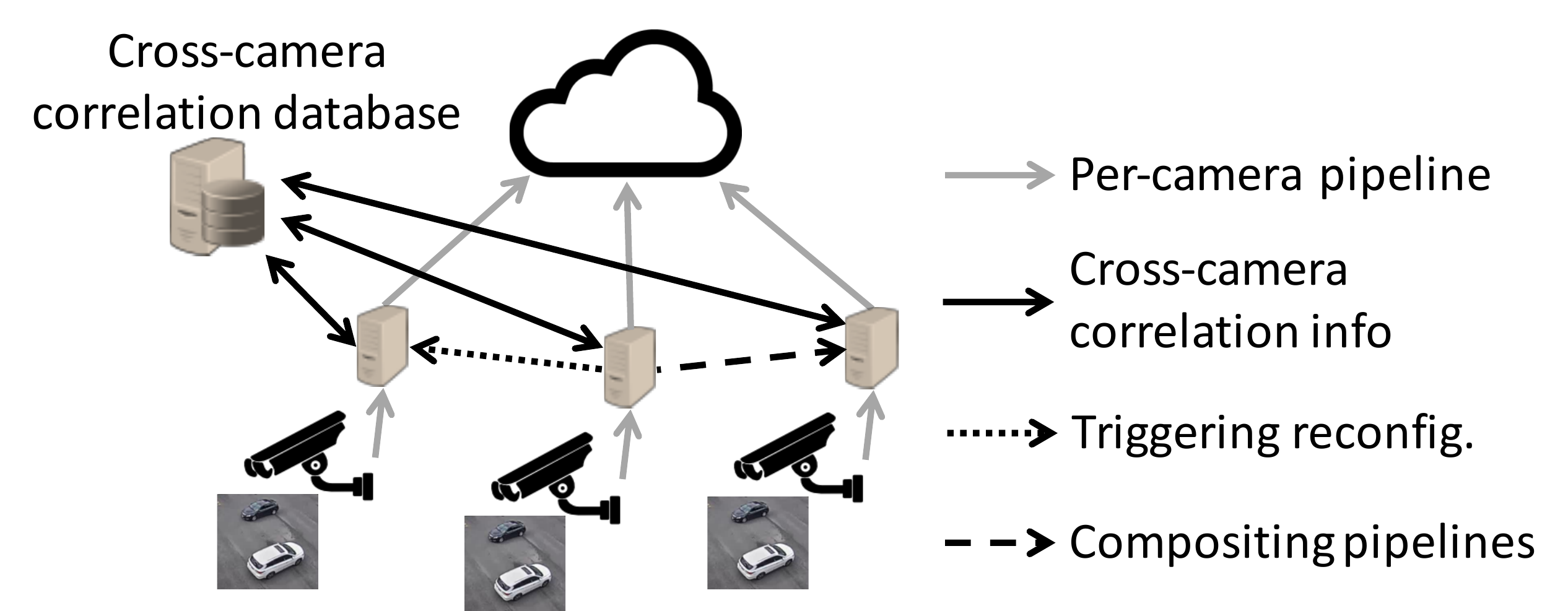}
    \vspace{-4mm}
    \tightcaption{End-to-end, cross-camera analytics architecture}
    \label{fig:new-architecture}
    \vspace{+2mm}
\end{figure}

\mypara{\#4: Fast analytics on stored video}
Recall from \S\ref{sec:trends} that spotlight search can involve tracking an object \textit{backward} for short periods of time to its first appearance in the camera network.
This requires a new feature, lacking in most video stream analytics systems: fast analysis of stored video data, {\em in parallel} with analytics on live video.
Stored video must be processed with very low latency (\eg several seconds), as subsequent tracking decisions depend on the results of the search.
In particular, this introduces a new requirement: processing many seconds or minutes of stored video at \textit{faster-than-real-time} rates.

\mypara{Putting it all together}
Figure~\ref{fig:new-architecture} depicts a new video analytics system that incorporates these proposed changes, along with two new required interfaces. 
Firstly, the correlation database must expose an interface to the analytics pipelines that reveals the spatio-temporal correlation between any two cameras.
Secondly, pipelines must support an interface for real-time communication, to (1) trigger inference (C1 in \S\ref{subsec:cost}) and (2) share inference results (A1 and A2 in \S\ref{subsec:accuracy}). 
This channel can be extended to support the sharing of resource availability (C2) and optimal configurations (C3).

\vspace{+2mm}
\subsection{Technical challenges}
\label{subsec:challenges}

In this section, we highlight the technical challenges that must be resolved to fully leverage cross-camera correlations.

\mypara{1) Learning cross-camera correlations}
To enable multi-camera optimizations, cross-camera correlations need to be extracted in the first place.
We envision two basic approaches. 
One is to rely on domain experts, \eg system administrators or developers who deploy cameras and models.
They can, for example, calibrate cameras to determine the overlapped field of view, based on camera locations and the floor plan. 
A data-driven approach is to {\em learn} the correlations from the inference results; \eg if two cameras identify the same person in a short time interval, they exhibit a {\em content correlation}.

The two approaches represent a tradeoff--- the data-driven approach can better adapt to dynamism in the network, but is more computationally expensive (\eg it requires running an offline, multi-person tracker \cite{ristani2018} on all video feeds to learn the correlations).
A hybrid approach is also possible: let domain experts establish the initial correlation database, and dynamically update it by periodically running the tracker.
This by itself is an interesting problem to pursue.

\mypara{2) Resource management in camera clusters}
Akin to clusters in the cloud, a set of cameras deployed by an enterprise also represents a ``cluster'' with compute capacity and network connectivity.
Video analytics work must be assigned to the different cameras in proportion to their available resources, while also ensuring high utilization and overall performance.
While cluster management frameworks~\cite{mesos} perform resource management, two differences stand out in our setting.
Firstly, video analytics focuses on analyzing video \textit{streams}, as opposed to the batch jobs \cite{spark} dominant in big data clusters.
Secondly, our spatio-temporal correlations enable us to \textit{predict} person trajectories, and by extension, forecast future resource availability, which adds a new, temporal dimension to resource management.

Networking is another important dimension. 
Cameras often need to share data in real-time (\eg A1, A2 in \S\ref{subsec:accuracy}, \#3 in \S\ref{subsec:missing}).
Given that the links connecting these cameras could be constrained wireless links, the network must also be appropriately scheduled jointly with the compute capacities.

Finally, given the long-term duration of video analytics jobs, it will often be necessary to {\em migrate} computation across cameras (\eg C2 in \S\ref{subsec:cost}, \#2 in \S\ref{subsec:missing}). Doing so will require considering both the overheads involved in transferring state, and in loading models onto the new camera's GPUs.

\mypara{3) Rewind processing of videos}
Rewind processing (\#4 in \S\ref{subsec:missing})---analyzing recently recorded videos---in parallel with live video requires careful system design.
A na\"ive solution is to ship the video to a cloud cluster, but this is too bandwidth-intensive to finish in near-realtime. 
Another approach is to process the video where it is stored, but a camera is unlikely to have the capacity to do this at faster-than-real-time rates, while also processing the live video.

Instead, we envision a MapReduce-like solution, which utilizes the resources of many cameras by (1) partitioning the video data and (2) calling on multiple cameras (and cloud servers) to perform rewind processing in parallel.
Care is required to orchestrate computation across different cameras, in light of their available resources (compute and network). Statistically, we expect rewind processing to involve only a small fraction of the cameras at any point in time, thus ensuring the requisite compute capacity.


\section{Related Work}

Finally, we put this paper into perspective by briefly surveying topics that are related to multi-camera video analytics.

\mypara{Video analytics pipelines}
Many systems today exploit a combination of camera, smartphone, edge cluster, and cloud resources to analyze video streams ~\cite{noscope,mcdnn,videostorm,chameleon,Liu2018}. Low cost model design~\cite{Fang2018,noscope,Shen2018}, partitioned processing ~\cite{Zhang2015,Hung2018,Chen2015}, efficient offline profiling~\cite{mcdnn,chameleon}, and compute/memory sharing~\cite{LiKamWa2015,Mathur2017,Mainstream2018} have been extensively explored.
Our goal, however, is to meet the joint objectives of high accuracy and cost efficiency in a multi-camera setting.
Focus \cite{focus} implements low-latency search, but targets historical video, and importantly does not leverage any cross-camera associations.
Chameleon \cite{chameleon} exploits content similarity across cameras to amortize query profiling costs, but still \textit{executes} video pipelines in isolation.
In general, techniques for optimizing individual video pipelines are orthogonal to a cross-camera analytics system, and could be co-deployed.

\mypara{Camera networks}
Multi-camera networks (\eg~\cite{aghajan2009multi,abas2014wireless,miller2015scmesh})
and applications (\eg~\cite{kamal2013information,dieber2011resource})
have been explored as a means to enable cross-camera communication (\eg over WiFi), and allow power-constrained cameras to work collaboratively.

Our work is built on these communication capabilities, but focuses on building a custom data analytics stack that spans a cluster of cameras.
While some camera networks do perform analytics on video feeds (\eg~\cite{song2010tracking, taj2011distributed,Zhang2015}), 
they have specific objectives (\eg minimizing bandwidth utilization), and fail to address the growing resource cost of video analytics, or provide a common interface to support various vision tasks.

\mypara{Geo-distributed data analytics}
Analyzing data stored in geo-distributed services (\eg data centers) is a related and well-studied topic (\eg~\cite{pu2015low,vulimiri2015wanalytics,Gaia2017}). 
The key difference in our setting is that camera data exhibits spatio-temporal correlations, which as we have seen, can be used to achieve major resource savings and improve analytics accuracy.

\section{Conclusions}

The increasing prevalence of enterprise camera deployments presents a critical opportunity to improve the efficiency and accuracy of video analytics via spatio-temporal correlations. The challenges posed by cross-camera applications call for a major redesign of the video analytics stack. We hope that the ideas in this paper both motivate this architectural shift, and highlight potential technical directions for its realization.

\vspace{+1mm}
\bibliographystyle{abbrv}
\begin{footnotesize}
\bibliography{hotnets17}
\end{footnotesize}

\end{document}